\documentclass[aip, pof,
 amsmath,
 amssymb,
reprint,
showpacs,
showkeys,
]{revtex4-1}
\usepackage{graphicx}
\pdfoutput=1
\usepackage{dcolumn}% Align table columns on decimal point
\usepackage{bm}% bold math
\begin{document}
%\preprint{AIP/123-QED}
\title{Self-similar shock dynamics satisfying the inviscid Burgers equation in planar, cylindrical and spherical problems}
\author{M. I. Radulescu}
   \email{matei@uottawa.ca}
   %\author{A. Sow}
\affiliation{Department of Mechanical Engineering, University of Ottawa, Ottawa (ON) K1N 6N5 Canada}
\date{\today}

\begin{abstract}
The solution of self-similar shock dynamics satisfying the inviscid Burgers equation are provided in closed form for planar, cylindrical and spherical problems.  The approach follows Lee's method for obtaining self-similar solutions for the Euler equations describing compressible fluid dynamics. Closed form solutions are provided for the two types of self-similar solutions: of the first kind, where the shock dynamics are constrained by integral relations, and the second kind, constrained by the internal requirement of the regularity of solutions along limiting characteristics.  The solutions obtained illustrate simply the theoretical underpinning of Taylor-Sedov blast waves (self-similarity of the first kind) and Guderley implosion problems (self-similarity of the second kind). 
\end{abstract}

\pacs{boh}% PACS, the Physics and Astronomy
                             % Classification Scheme.
\maketitle

\section{Introduction}
The inviscid Burgers' equation \cite{burgers1948} for a scalar $c$, namely,
\begin{align}
\frac{\partial c}{\partial t}+c\frac{\partial c}{\partial x}=0
\end{align}
is a single-sided non-linear wave equation which serves as a prototype for studying problems of compressible flows and fluid dynamics alike, since it contains the convective non-linearity of the Euler equations of inviscid flow. The convection speed  $c$ in front of the space derivative being the same as the transported variable inside the time and space partial derivatives, gives rise to wave steepening in compressive regions leading to shock formation.  

There are a wide range of problems where the Euler equations degenerate into Burgers equation describing their dynamics.  In gas dynamics, weak shocks are described by the inviscid Burgers equation  \cite{chandrasekhar1943, whithambook}.  Reactive compressible flow problems also often degenerate into reactive forms of the inviscid Burgers equation \cite{bdzil_stewart_1986, clavin2002, faria_kasimov_rosales_2015}.  Ad hoc extensions of the inviscid Burgers equation have also been proposed as toy mathematical models to explore certain non-linear aspects of the dynamics of compressible reactive flows \cite{Fickett1979, radulescu2011nonlinear, kasimov2013model, mi2015influence}.  

In the present communication, we wish to derive the similarity solutions for shock waves satisfying the inviscid Burgers equation.  They will be the analogue of the well known Taylor-Sedov \cite{taylor1950, sedov1959} point blast explosion problems of compressible flows and the Guderley solution for self-similar shock implosion \cite{guderley1942}.  We follow the systematic method used by Lee to unify such problems of self-similar shock propagation in perfect gases \cite{lee}.  To the best of our knowledge, only explosion solutions have been proposed by Mi et al. for the planar problem modelled by the inviscid Burgers equation, using the method of characteristics \cite{mi2015influence}.

\section{The Burgers-Fickett equation}
The general continuity equation is
\begin{align}
\frac{\partial \rho}{\partial t}+\frac{\partial q}{\partial x}&=-q \frac{j}{x}
\end{align}
with $j$=0, 1 or 2 for planar, cylindrical and spherical geometries respectively.  The density of the quantity being conserved is $\rho$ and $q$ is its flux (transfer rate of the quantity conserved per unit time per unit area).  
 
Following Fickett \cite{Fickett1979, Fickett1985}, the simplest model for describing gas dynamics in a toy model is to model the flux $q$ as a function of $\rho$ alone.  The simplest model for the flux is the Burgers flux given by $\frac{\rho^2}{2}$.  The resulting non-linear wave equation becomes the Burgers-like equation
\begin{equation}
\frac{\partial \rho}{\partial t}+\rho \frac{\partial \rho}{\partial x}=-\frac{\rho^2}{2} \frac{j}{x} \label{eq:burgers}
\end{equation} 
We present similarity solutions to this model equation. This also serves to illustrate the solution methodology that can be used for extensions to this model for other flux functions in kinematic waves of material flow (e.g., traffic flow, pedestrian flow, flood waves, etc....) as pioneered by Lighthill and Whitham \cite{lighthill1955a, lighthill1955b}.  

\section{Dynamics of Burgers' shocks in similarity variables}
We are concerned with problems of shock dynamics.  We will consider the problem of strong shocks, where $\rho=0$ ahead of the shock.  Let $X(t)$ denote the shock's trajectory and $\dot{X}(t)$ its speed. We introduce the similarity variables 
\begin{align}
\zeta = \frac{x}{X(t)}\\ \label{eq:zeta}
\phi=\frac{\rho}{\dot{X}(t)}
\end{align}
We change variables from $(x, t)$ to $(\zeta(x,t) = \frac{x}{X(t)}, t'(t)=t)$.  With this change of variables, partial derivatives become:
\begin{align}
\left(\frac{\partial}{\partial t}\right)_x&=  \left(\frac{\partial }{\partial \zeta}\right)_{t'} \left(\frac{\partial \zeta}{\partial t}\right)_x+\left(\frac{\partial }{\partial t'}\right)_{\zeta} \left(\frac{\partial t'}{\partial t}\right)_x\\
&=-\zeta\frac{\dot{X}}{X}\left(\frac{\partial }{\partial \zeta}\right)_{t'}+\left(\frac{\partial }{\partial t'}\right)_{\zeta}
\end{align}
and 
\begin{align}
\left(\frac{\partial}{\partial x}\right)_t =  \left(\frac{\partial }{\partial \zeta}\right)_{t'} \left(\frac{\partial \zeta}{\partial x}\right)_t =\frac{1}{X}\left(\frac{\partial }{\partial \zeta}\right)_{t'}
\end{align}
With these transformations and seeking solutions where $\phi=\phi(\zeta)$ only, \eqref{eq:burgers} can be re-written as an ordinary differential equation:
\begin{equation}
\dot{\phi}=-\frac{\frac{j}{2}\frac{\phi^2}{\zeta}+\theta \phi}{\phi-\zeta} \label{eq:self-similar0}
\end{equation} 
provided 
\begin{equation}
\theta \equiv \frac{\ddot{X}(t)X(t)}{\dot{X}(t)^2} \label{eq:self-similar}
\end{equation}
is not a function of $t$.  Since it is not a function of $\zeta$ by construction, it can only be a constant.  We have thus reduced the problem to determining $\theta$ and the distribution $\phi(\zeta)$ given by \eqref{eq:self-similar0}.  

Since we can re-write \eqref{eq:self-similar} as:
\begin{align}
\theta=\frac{\ddot{X}X}{\dot{X}^2}=\frac{\ddot{X}/\dot{X}}{\dot{X}/X}=\frac{\mathrm{\ln \dot{X}}}{\mathrm{\ln X}}= \mathit{constant}
\end{align} 
we obtain immediately that the shock speed depends on distance as a power law:
\begin{align}
\dot{X}=B X^\theta \label{eq:power}
\end{align}
where $B$ is an integration constant. 

The shock decay coefficient $\theta$ is problem dependent.  Below, we illustrate how to obtain $\theta$ in two types of self-similar problems of the Burgers equation: for similarity solutions of the first and second kind.  We draw from blast wave theory of gas dynamics \cite{lee}.

\section{Similarity solutions of the first kind: shock motion given release of material at a point source}
The first problem we wish to solve is the analogous problem to the Taylor-Sedov problem of blast waves, where the energy addition in a gas at a point, along a line or along a plane gives rise to a decaying strong shock followed by a self-similar profile of the field variables (speed, density and pressure).  For the Burgers equation, the analogous problem is thus the sudden release of fixed material $m_j$ in a medium initially at $\rho=0$. $m_1$ is the amount of material per unit area in the planar problem, $m_2$ the material per unit length for the cylindrical problem and $m_3$ the material for the spherical problem.  We wish to determine the law of shock decay and the self-similar distribution of density behind the shock $\rho(x,t)$ satisfying \eqref{eq:burgers} for the planar, cylindrical and spherical problems.

The point, line or plane release problems posed requires the material engulfed by the decaying shock to remain constant, since $\rho=0$ ahead of the shock, i.e., 
\begin{align}
\int_0^{X(t)} \rho k_j x^j \mathrm{d}x = m_j
\end{align}
where $m_j$ is a constant. The geometrical factor $k_j$ takes values $4\pi$, $2\pi$, and $1$ for spherical, cylindrical and planar waves.  In terms of the similarity variables, and using \eqref{eq:power}, we obtain 
\begin{align}
 m_j= B X(t)^{\theta+j+1} I_j
\end{align}
where 
\begin{align}
I_j=\int_0^{1} \phi k_j \zeta^j \mathrm{d}\zeta \label{eq:I}
\end{align}
Since $m_j$ is not a function of $t$, the exponent of $X(t)$ needs to vanish for a self-similar solution. This requires 
\begin{align}
\theta=-(1+j)
\end{align}   
The blast decay coefficient $\theta$ takes values of -1, -2 and -3 for planar, cylindrical and spherical problems respectively.  We also now relate the mass deposition to the integral $I_j$:
\begin{align}
m_j = B I_j
\end{align}

Note that the mass integral permitted us to determine the shock decay profile for the kinematic wave considered.  In the  Taylor-Sedov solution, the integral determining the blast solution is the energy integral of the total energy engulfed by the shock, made equal to the energy deposition. 

For decaying shock waves from $X(t=0) = 0$, \eqref{eq:power} can be further integrated to obtain
\begin{equation}
X=C t^N
\end{equation}
with the constants $C=\left(B(1-\theta)\right)^N$ and $N=\frac{1}{1-\theta}$.  Spherical blast wave radius $X$ grows like $t^{1/4}$, cylindrical waves grow like $t^{1/3}$ and planar waves grow like $t^{1/2}$.

The speed of the shock satisfies the shock jump condition:
\begin{equation}
\dot{X}=\frac{[\frac{1}{2}\rho^2 ]_0^s}{[\rho]_0^s}=\frac{1}{2}\rho
\end{equation}
where $[\alpha]_0^s=\alpha_s-\alpha_0$, such that 
\begin{equation}
\phi(\zeta=1) = 2
\end{equation}
With this boundary condition and the blast decay coefficient $\theta$ known, the self-similar profile $\phi(\zeta)$ satisfying \eqref{eq:self-similar0} is readily found.  By inspection, the solution is
\begin{equation}
\phi=2 \zeta
\end{equation}
for the planar, cylindrical and spherical geometries alike, as it satisfies \eqref{eq:self-similar0} identically.  

With $\phi(\zeta)$ known, the constant $I_j$ given by \eqref{eq:I} can readily be found
\begin{align}
I_j=\frac{2 k_j}{j+2}
\end{align}
It takes values of 1, $4/3\pi$ and $2\pi$ for planar, cylindrical and spherical waves respectively.  The constant $B$ in the self-similar solution becomes $m_1$, $\frac{3m_2}{4 \pi}$ and $\frac{m_3}{2 \pi}$ for planar, cylindrical and spherical waves respectively. 
The self-similar shock motion given by \eqref{eq:power} becomes
\begin{align}
\mathrm{planar:}\; & \dot{X}=m_1 X^{-1} \\
\mathrm{cylindrical:}\: & \dot{X}=\left( \frac{3m_2}{4 \pi} \right) X^{-2}  \\
\mathrm{spherical:}\; & \dot{X}=\left( \frac{m_3}{2 \pi} \right) X^{-3} 
\end{align}

\begin{figure}
	\center
	\includegraphics[width=\columnwidth]{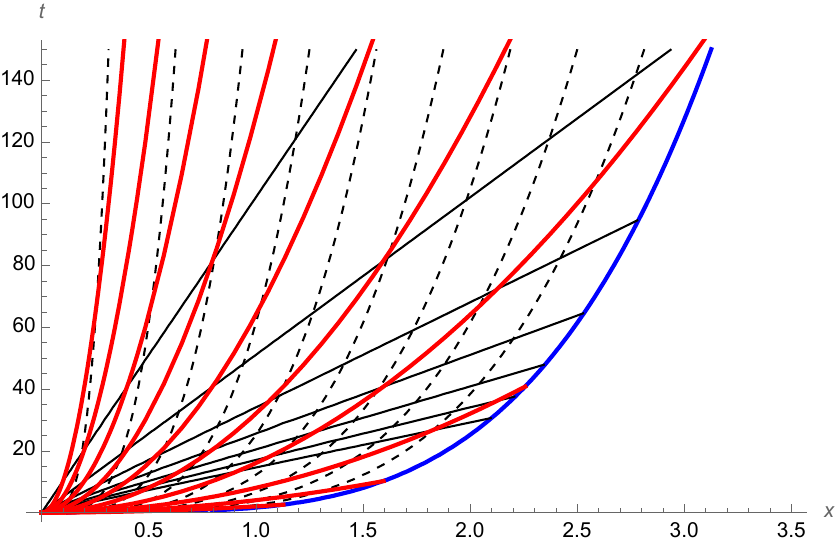}
	\caption{Space-time diagram of the spherical explosion problem, with $m=1$; blue line is the shock trajectory, black solid lines are contours of $\rho$, black dashed lines are contours of the similarity variable $\zeta$ and red lines are characteristics.}
	\label{fig:explosion}
\end{figure} 

Fig.\ \ref{fig:explosion} illustrates the solution for the spherical explosion problem.  Interestingly, the interior solution is not a function of the initial mass addition, and is given by $\rho=\frac{x}{2t}$; only the shock trajectory is.  The trajectory of characteristics is also shown; these are simple power laws by integrating $\mathrm{d}x/\mathrm{d}t=\rho$.  The region bounded by the shock is the domain of dependence of the lead shock motion, with characteristics catching up to the lead shock. The planar and cylindrical problems behave similarly.  The solution obtained for the planar problem agrees with the one provided by Mi et al.\ \cite{mi2015influence}.

\section{Similarity solutions of the second kind: self-similar shock implosion}
The second problem we will solve is one in which the shock decay coefficient $\theta$ cannot be obtained from an integral relation of conservation of mass, momentum or energy. In blast wave theory \cite{lee}, these are said to be similarity solutions of the second kind. We now solve the problem of shock implosion for the cylindrical and spherical problems, i.e., the Guderley problem of shock dynamics \cite{guderley1942}.  

Self-similarity still requires power law solutions for the shock dynamics given by \eqref{eq:power} with the distribution behind the shock given by \eqref{eq:self-similar0}. Inspection of \eqref{eq:self-similar0} suggests that the profiles will be singular when the denominator vanishes, i.e, when $\zeta=\phi$.  A regular solution without blow-up corresponds to requiring that the numerator of \eqref{eq:self-similar0} vanishes simultaneously at this point. Denoting the internal point when this is satisfied by a *, we have $\zeta^*=\phi^*$.  Setting the numerator to zero yields immediately
\begin{equation}
\theta=-j/2
\end{equation} 
The profiles can now be obtained by direct integration of the ODE's \eqref{eq:self-similar0}.  With the appropriate boundary condition at the shock $\phi(1)=2$, one obtains the wave structure solution:  
\begin{align}
\phi=\frac{2}{\zeta}\; \mathrm{for}\: j=2 \mathrm{(spherical)}\\
\phi=\frac{2}{\sqrt{\zeta}} \; \mathrm{for}\; j=1 \mathrm{(cylindrical)}
\end{align}
The saddle point in the solution lies at $\zeta^* = 2^{1/2}$ for spherically imploding waves and at $\zeta^* = 2^{3/2}$ for cylindrically imploding waves.

While the solution is complete, it is worth further elucidating the meaning of the regularization criterion used to obtain the solution and its link to the underlying wave motion dynamics of Burgers' equation.  Burgers equation \eqref{eq:burgers} is a hyperbolic equation with a single family of characteristics curves given 
\begin{align}
\frac{\mathrm{d}x}{\mathrm{d}t}=\rho
\end{align}
which can be re-written as:
\begin{align}
\frac{\mathrm{d}x}{\mathrm{d}t}=\phi \dot{X} \label{eq:dxdt1}
\end{align}

A line of $\zeta=\mathit{constant}$ corresponds to 
\begin{equation}
\mathrm{d}\zeta(x,t)=\left(\frac{\partial \zeta}{\partial x}\right)_t\mathrm{d}x+\left(\frac{\partial \zeta}{\partial t}\right)_x\mathrm{d}t=0
\end{equation}
i.e., 
\begin{equation}
\frac{\mathrm{d}x}{\mathrm{d}t}=-\frac{\left(\frac{\partial \zeta}{\partial t}\right)_x}{\left(\frac{\partial \zeta}{\partial x}\right)_t}
\end{equation}
Evaluating the partial derivatives from \eqref{eq:zeta}, the last equation can be re-written as: 
\begin{equation}
\frac{\mathrm{d}x}{\mathrm{d}t}=\zeta \dot{X} \label{eq:dxdt2}
\end{equation}
By comparing \eqref{eq:dxdt1} and \eqref{eq:dxdt2}, this signifies that the locus of $\zeta=\phi$ corresponds to the location where a characteristic surface coincides with a constant $\zeta$ line.  This is the limiting characteristic, serving as an event horizon for the dynamics of the imploding shock wave.  Along the limiting characteristic, $\phi$ remains constant by virtue of the right hand side of  \eqref{eq:self-similar0} vanishing, this requires the nominator to vanish.  Within the structure of the imploding shock wave, non-linearity leads to wave compression waves according to Burgers eauation, amplified by geometrical terms involving $j$ in \eqref{eq:self-similar0}.  The situation is thus perfectly analogous to the gas-dynamics problem of Guderley. 

\begin{figure}
	\center
	\includegraphics[width=\columnwidth]{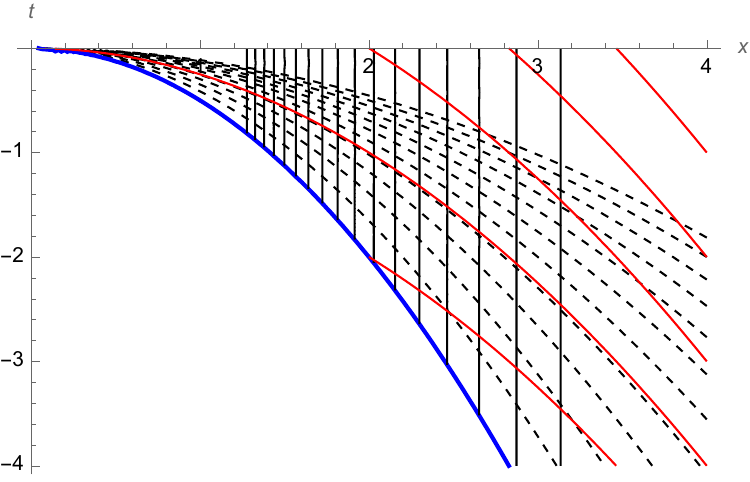}
	\caption{Space-time diagram of the spherical implosion problem, with $B=-1$; blue line is the shock trajectory, black solid lines are contours of $\rho$, black dashed lines are contours of the similarity variable $\zeta$ and red lines are characteristics.}
	\label{fig:implosion}
\end{figure}

Interestingly, the solutions for the implosion problem leads to interior solutions independent of time, bounded by time-dependent shock motion and by the time-dependent limiting characteristic.  Fig.\ \ref{fig:implosion} shows the solution for the spherical implosion problem.  The density field is given by $\rho=2B/x$.  The trajectory of characteristics are given by simple power laws by integrating $\mathrm{d}x/\mathrm{d}t=\rho$.  The limiting characteristic is the one coincident with the $\zeta=2^{3/2}$ curve.  This separates the characteristics affecting the implosion dynamics from the ones not affecting it. The behaviour is thus perfectly analogous to the Guderley solution of gas dynamics.  

The cylindrical implosion problem is qualitatively similar.  The density field is given by $\rho=2B/\sqrt{x}$. 

\section{Conclusions}
We have derived closed form solutions form self-similar shock dynamics satisfying the inviscid Burgers equation.  Self-similar shock dynamics of the first kind can be obtained by integral relations, while the second kind requires the requirement of analyticity on the limiting characteristic.  We recover perfectly analogous results to those of the inviscid Euler equations, generalizing the Taylor-Sedov blast wave problem  (first kind) and Guderley problem of shock implosion (second kind) in a transparent manner.  The method can be extended to other one-way non-linear wave equations modelling kinematic waves in a straightforward way.  

\section*{Acknowledgements}
This work was supported by AFOSR grant FA9550-23-1-0214, with Dr.\ Chiping Li as program monitor.  The author also acknowledges financial support from NSERC Discovery Grant "Predictability of detonation wave dynamics in gases: experiment and model development". 

\bibliography{references}

\end{document}